\documentclass[prb,showpacs,preprintnumbers,amsfonts,amssymb,amsmath,floats,twocolumn,aps]{revtex4}

\usepackage{graphicx}
\usepackage{dcolumn}
\usepackage{bm}
\usepackage[final,dvips]{epsfig}
\usepackage{bm}
\usepackage{color}

\newcommand{\be}{\begin{equation}}
\newcommand{\ee}{\end{equation}}
\newcommand{\ba}{\begin{eqnarray}}
\newcommand{\ea}{\end{eqnarray}}
\newcommand{\ff}[1]{{\bm #1}}
\newcommand{\tr}{\mbox{tr}}
\newcommand{\Tr}{\mbox{Tr}}
\newcommand{\refeq}[1]{Eq.\ (\ref{eq:#1})}
\newcommand{\labeq}[1]{\label{eq:#1}}

\begin{document} 
  
\title[Non-perturbative construction of the Luttinger-Ward functional]
      {Non-perturbative construction of the Luttinger-Ward functional} 

\author{Michael Potthoff}

\affiliation{
Institut f\"ur Theoretische Physik und Astrophysik, 
Universit\"at W\"urzburg, Germany
}
 
\begin{abstract}
For a system of correlated electrons,
the Luttinger-Ward functional provides a link between static thermodynamic
quantities on the one hand and single-particle excitations on the other.
The functional is useful to derive several general properties of the system 
as well as for the formulation of thermodynamically consistent approximations. 
Its original construction, however, is perturbative as it is based on the 
weak-coupling skeleton-diagram expansion. Here, it is shown that the Luttinger-Ward 
functional can be derived within a general functional-integral approach. 
This alternative and non-perturbative approach stresses the fact that the 
Luttinger-Ward functional is universal for a large class of models.
\end{abstract} 
 

\maketitle 

\section{Introduction}

For a system of correlated electrons in equilibrium, there are several
relations \cite{AGD64,FW71,NO88} between static quantities which describe 
the thermodynamics of the system and dynamic quantities which describe 
its one-particle excitations.
Static quantities are given by the grand potential $\Omega$ 
and its derivatives with respect to temperature $T$, chemical potential 
$\mu$ etc.
The one-electron Green's function $\ff G = \ff G (i\omega_n)$ or the self-energy 
$\ff \Sigma = \ff \Sigma(i\omega_n)$,
on the other hand, are dynamic quantities which yield (equivalent) information 
on an idealized (photoemission or inverse photoemission) excitation process.

The Luttinger-Ward functional $\widehat{\Phi}[\ff G]$ provides a special 
relation between static and dynamic quantities with several important 
properties: \cite{LW60} 
First, the grand potential is obtained from the Luttinger-Ward functional
evaluated at the exact Green's function, $\Phi = \widehat{\Phi}[\ff G]$, via
\be
  \Omega = \Phi + \Tr \ln \ff G - \Tr \, \ff \Sigma \ff G \: .
\label{eq:om}
\ee
Second, the functional derivative of the Luttinger-Ward functional,
\be
  \frac{1}{T} \frac{\delta \widehat{\Phi}[\ff G]}{\delta \ff G} 
  = \widehat{\ff \Sigma}[\ff G] \: ,
\label{eq:der}
\ee
defines a functional $\widehat{\ff \Sigma}[\ff G]$ which gives the exact self-energy
of the system if evaluated at the exact Green's function.
The relation $\ff \Sigma = \widehat{\ff \Sigma}[\ff G]$ is independent from the 
Dyson equation $\ff G^{-1} = \ff G_0^{-1} - \ff \Sigma$.
Third, in the non-interacting limit:
\be
  \widehat{\Phi}[\ff G] \equiv 0 \quad \mbox{for} \: \ff U = 0 \; .
\label{eq:bl}  
\ee
Finally, the functional dependence $\widehat{\Phi}[\ff G]$ is completely determined
by the interaction part of the Hamiltonian and independent from the 
one-particle part:
\be
 \widehat{\Phi}[\ff G] \quad \mbox{universal} \: .
\label{eq:univ} 
\ee
This universality property can also be expressed as follows:
Two systems with the same interaction $\ff U$ but different one-particle 
parameters $\ff t$ (on-site energies and hopping integrals)
in the respective Hamiltonian are described by the same
Luttinger-Ward functional. 
Using Eq.\ (\ref{eq:der}), this implies that the functional 
$\widehat{\ff \Sigma}[\ff G]$ is universal, too.

\begin{figure}[b]
  \includegraphics[width=0.65\columnwidth]{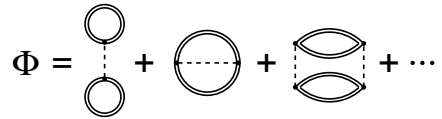}
\caption{
Classical definition of the Luttinger-Ward functional $\widehat{\Phi}[\ff G]$.
Double lines: fully interacting propagator $\ff G$. 
Dashed lines: interaction $\ff U$.
}
\label{fig:lw}
\end{figure}

If Ref.\ \onlinecite{LW60} it is shown by Luttinger and Ward that  
$\widehat{\Phi}[\ff G]$ can be constructed order by order in diagrammatic weak-coupling 
perturbation theory. 
$\Phi$ is obtained as the limit of the infinite series of closed diagrams without 
any self-energy insertions and with all free propagators in a diagram replaced by 
fully interacting ones (see Fig.\ \ref{fig:lw}). 
Generally, this skeleton-diagram expansion cannot be summed up to get a closed
form for $\widehat{\Phi}[\ff G]$. 
So, unfortunately, the explicit functional dependence $\widehat{\Phi}[\ff G]$ is 
actually unknown -- even for the most simple Hamiltonians such as the Hubbard model. 
\cite{Hub63}
The defining properties, Eqs.\ (\ref{eq:om}--\ref{eq:univ}), however, are easily verified.
\cite{LW60}

The Luttinger-Ward functional is useful for several general considerations:
With the help of $\widehat{\Phi}[\ff G]$ and the Dyson equation, the grand 
potential can be considered as a functional of the Green's function
$\Omega = \widehat{\Omega}[\ff G]$ or as functional of the self-energy
$\Omega = \widehat{\Omega}[\ff \Sigma]$, such that $\Omega$ is stationary 
at the physical $\ff G$ or $\ff \Sigma$. \cite{LW60,Bay62}
This represents a remarkable variational principle which connects static with
dynamic physical quantities.
The Luttinger-Ward functional is also used in the microscopic derivation of
some zero- or low-temperature properties of Fermi liquids as discussed in
Refs.\ \onlinecite{LW60,Lut60}.
The derivative of the functional, Eq.\ (\ref{eq:der}), shows the self-energy
to be gradient field when considered as a functional of the Green's function, 
$\widehat{\ff \Sigma}[\ff G]$. 
This fact is related to certain symmetry properties of two-particle Green's 
functions as originally noted by Baym and Kadanoff. \cite{BK61}
Furthermore, the Luttinger-Ward functional is of great importance in the 
construction of thermodynamically consistent approximations. 
So-called conserving approximations virtually start from the Luttinger-Ward
functional. \cite{BK61,Bay62}
This is essential to prove these approximations to respect a number of 
macroscopic conservation laws. 
The Hartree-Fock and the random-phase approximations are well-known examples.
These ``classical'' conserving approximations are essentially limited to the
weak-coupling regime.
However, the Luttinger-Ward functional can also be used to construct 
non-perturbative approximations.
This was first realized in the context of the dynamical mean-field theory 
(DMFT) for lattice models of correlated electrons. \cite{MV89,GK92a,Jar92,GKKR96}
Here, one exploits the universality of the functional, Eq.\ (\ref{eq:univ}),
to achieve an (approximate) mapping of the original lattice model onto a 
simpler impurity model with the same interaction part.
The fact that $\widehat{\Phi}[\ff G]$ is the same for a large class of 
systems, has recently been shown \cite{Pot03a,Pot03b} to be the key feature 
that allows to construct several non-perturbative and 
thermodynamically consistent approximations. \cite{PAD03,DAH+04}
This idea has been termed ``self-energy-functional approach'' (SFA).

Such general considerations remain valid as long as the Luttinger-Ward functional 
is well defined.
This presupposes that the skeleton-diagram expansion is convergent or at least
that formal manipulations of diagrammatic quantities are consistent in themselves
and eventually lead to physically meaningful results.
Provided that one can assure that no singular point is passed when starting from 
the non-interacting Fermi gas and increasing the interaction strength, this seems 
to be plausible.
A strict proof that the skeleton-diagram expansion is well-behaved, however, will 
hardly be possible in most concrete situations.
On the contrary, it is well known that the expansion is questionable in a number 
of cases, e.g.\ in case of a symmetry-broken state or a state that is not 
``adiabatically connected'' to the non-interacting limit, such as a Mott insulator. 
The skeleton-diagram expansion may break down even in the absence of any spontaneous 
symmetry breaking in a (strongly correlated) state that gradually evolves from a 
metallic Fermi liquid.
This has explicitly been shown by Hofstetter and Kehrein \cite{HK99} for the 
narrow-band limit of the single-impurity Anderson model 
(see Refs.\ \onlinecite{Keh98,Kot99} for a discussion of possible physical consequences).
Generally speaking there is no strict argument available that ensures the convergence 
of the skeleton-diagram expansion in the strong-coupling regime.

The purpose of the present paper is to show that a construction of the 
Luttinger-Ward functional is possible that does not make use of the  
skeleton-diagram expansion.
The proposed construction is based on a standard 
functional-integral approach and avoids the formal complications 
mentioned above.
Thereby, one achieves an alternative and in particular non-perturbative route
to the general properties of correlated electron systems derived from the functional, 
to the dynamical mean-field theory as well as to the self-energy-functional approach.
It should be stressed that the intended construction of the Luttinger-Ward 
functional requires more than a simple definition of the quantity $\Phi$
(which could trivially be achieved by using Eq.\ (\ref{eq:om}): 
$\Phi \equiv \Omega - \Tr \ln \ff G + \Tr \, \ff \Sigma \ff G$).
The task is rather to provide a functional $\widehat{\Phi}[\ff G]$ with the 
properties Eqs.\ (\ref{eq:om}--\ref{eq:univ}).

Previous approaches are either perturbative or cannot prove 
Eqs.\ (\ref{eq:om}--\ref{eq:univ}):
A construction of the Luttinger-Ward functional different from the original 
one \cite{LW60} has been given by Baym: \cite{Bay62}
The existence of $\widehat{\Phi}[\ff G]$ is deduced from a ``vanishing curl
condition'', 
$\delta \Sigma(1,1') / \delta G(2',2) = \delta \Sigma(2,2') / \delta G(1',1)$,
which is derived from an analysis of the functional dependence of $\ff G$ on
an arbitrary (time-dependent) external perturbation $\ff J$.
However, an independent functional relation $\ff \Sigma = \widehat{\ff \Sigma}[\ff G]$ 
is required in addition. 
In Ref.\ \onlinecite{Bay62} the latter is assumed to be given by the (full or 
by a truncated) skeleton-diagram expansion, and consequently this approach 
is perturbative again.

As also shown in Ref.\ \onlinecite{Bay62}, the Green's function in the presence 
of an external field $\ff J$ can be derived from the grand potential 
$\widehat{\Omega}'[\ff J]$ 
as $\widehat{\ff G}(\ff J) = (1/T) \delta \widehat{\Omega}'[\ff J] / \delta \ff J$.
Using the inverse functional, $\widehat{\ff J}[\ff G]$, Legendre transformation 
yields 
$\widehat{\Omega}[\ff G] = \widehat{\Omega}'[\widehat{\ff J}[\ff G]] - \Tr \,\ff G \, 
\widehat{\ff J}[\ff G]$.
This (non-perturbative) functional and the Dyson equation can be used to define
$\widehat{\Phi}[\ff G] \equiv \widehat{\Omega}[\ff G] - \Tr \ln \ff G + \Tr \, 
(\ff G_0^{-1} - \ff G^{-1} )\ff G$.
This idea is in the spirit of the effective action approach. \cite{SK01,SK03,Geo04}
Here, however, 
the problem is that the universality of $\widehat{\Phi}[\ff G]$, Eq.\ (\ref{eq:univ}),
cannot be proven.
The Luttinger-Ward functional constructed in this way explicitly depends on $\ff G_0$ 
and thus on the one-particle parameters $\ff t$.

The paper is organized as follows:
The next section briefly introduces the notations and the quantities of interest.
The construction of the Luttinger-Ward functional is described in Sec.\ \ref{sec:lw}.
Sec.\ \ref{sec:dis} gives a brief discussion of the properties of the functional
and its use within the dynamical mean-field theory and the self-energy-functional
approach.
The results are summed up in Sec.\ \ref{sec:sum}.

\section{Static and dynamic quantities}
\label{sec:sd}

Consider a system of electrons at temperature $T$ and chemical potential $\mu$ 
in thermal equilibrium and let $H = H(\ff t, \ff U) = H_0(\ff t) + H_{1}(\ff U)$
be its Hamiltonian where
\ba
  H_0(\ff t) &=& \sum_{\alpha\beta} t_{\alpha\beta} \:
  c^\dagger_\alpha c_\beta \; ,
\nonumber \\  
  H_{1}(\ff U) &=& \frac{1}{2} 
  \sum_{\alpha\beta\gamma\delta} U_{\alpha\beta\delta\gamma} \:
  c^\dagger_\alpha c^\dagger_\beta c_\gamma c_\delta \: .
\ea
An index $\alpha$ refers to an arbitrary set of quantum numbers characterizing
a one-particle basis state.
If $N$ is the total particle-number operator, the grand potential of the system 
is given by $\Omega_{\ff t,\ff U} = - T \ln Z_{\ff t,\ff U}$ where
\be
  Z_{\ff t,\ff U} = \tr \exp(-(H(\ff t, \ff U) - \mu N)/T) 
\ee
is the partition function. Here and in the following the dependence of all
quantities on the one-particle parameters $\ff t$ and the interaction
parameters $\ff U$ is made explicit through the subscripts.

Using a matrix notation, the free one-particle Green's function is denoted by 
$\ff G_{\ff t, 0}$. 
Its elements (for fixed $\mu$) are given by:
\be
  G_{\ff t, 0, \alpha\beta}(i\omega_n) = 
  \left(\frac{1}{i\omega_n + \mu - \ff t}\right)_{\alpha\beta} \: .
\ee
Here $i \omega_n = i(2n+1)\pi T$ is the $n$-th Matsubara frequency.
The fully interacting Green's function is denoted by $\ff G_{\ff t, \ff U}$.
Using Grassmann variables 
$\xi_\alpha(i\omega_n)= T^{1/2} \int_0^{1/T} d\tau \: e^{i\omega_n\tau} \xi_\alpha(\tau)$ 
and $\xi^\ast_\alpha(i\omega_n)= T^{1/2} \int_0^{1/T} d\tau \: e^{-i\omega_n\tau} \xi^\ast_\alpha(\tau)$,
its elements can be written as \cite{NO88}
\ba
&&
  G_{\ff t, \ff U, \alpha\beta}(i\omega_n) = 
  - \langle \xi_\alpha(i\omega_n) \xi^\ast_\beta(i\omega_n)
  \rangle_{\ff t, \ff U}
\nonumber \\ && = \frac{-1}{Z_{\ff t,\ff U}}
  \int D \xi D \xi^\ast 
  \xi_\alpha(i\omega_n) \xi^\ast_\beta(i\omega_n)
  \exp\left( A_{\ff t, \ff U,\xi\xi^\ast} \right)
\nonumber \\
\labeq{gfunct}
\ea
where
\ba
   &&
   A_{\ff t, \ff U,\xi\xi^\ast} =
   \sum_{n,\alpha\beta} \xi_\alpha^\ast(i\omega_n) 
   ((i\omega_n + \mu)\delta_{\alpha\beta} - t_{\alpha\beta})
   \xi_\beta(i\omega_n) 
\nonumber \\   
   &&- 
   \frac{1}{2} \sum_{\alpha\beta\gamma\delta} U_{\alpha\beta\delta\gamma} \int_0^{1/T} 
   \!\! d\tau \:
   \xi_\alpha^\ast(\tau) 
   \xi_\beta^\ast(\tau)
   \xi_\gamma(\tau) 
   \xi_\delta(\tau) 
\ea
is the action.
Finally, the self-energy is defined as
\be
  \ff \Sigma_{\ff t, \ff U} = 
  \ff G_{\ff t, 0}^{-1} -
  \ff G_{\ff t, \ff U}^{-1} \: .
\labeq{dyson}
\ee

The goal is to construct a functional $\widehat{\Phi}_{\ff U}[\ff G]$ (where $\ff G$ 
is considered as a free variable) which vanishes in the non-interacting case, 
$\widehat{\Phi}_{0}[\ff G] = 0$ [Eq.\ (\ref{eq:bl})], which is universal, i.e.\
independent of $\ff t$ [Eq.\ (\ref{eq:univ})], which yields 
$\widehat{\Phi}_{\ff U}[\ff G_{\ff t, \ff U}] = \Omega_{\ff t, \ff U} - 
\Tr \ln \ff G_{\ff t, \ff U} + \Tr ( \ff \Sigma_{\ff t, \ff U} \ff G_{\ff t, \ff U} )$
if evaluated at the exact Green's function $\ff G = \ff G_{\ff t, \ff U}$ 
[Eq.\ (\ref{eq:om})], 
and the derivative of which is a functional
$\widehat{\ff \Sigma}[\ff G]$ with $\widehat{\ff \Sigma}[\ff G_{\ff t, \ff U}] = 
\ff \Sigma_{\ff t, \ff U}$ [Eq.\ (\ref{eq:der})].
(Here the notation $\Tr \ff A \equiv T \sum_n \sum_\alpha e^{i\omega_n0^+} 
A_{\alpha\alpha}(i\omega_n)$ is used.
$0^+$ is a positive infinitesimal.
Functionals $\widehat{A} = \widehat{A}[\cdots]$ are indicated by a hat and
should be distinguished clearly from physical quantities $A$.)
 
For the classical construction of $\widehat{\Phi}_{\ff U}[\ff G]$ via the skeleton-diagram
expansion (Fig.\ \ref{fig:lw}), these properties are easily verified:
The universality of the functional [Eq.\ (\ref{eq:univ})] is obvious as any diagram depends 
on $\ff U$ and on $\ff G$ {\em only}; there is no explicit dependence on the free Green's 
function $\ff G_{\ff t, 0}$, i.e.\ no explicit dependence on $\ff t$.
Since there is no zeroth-order diagram, $\widehat{\Phi}_{\ff U}[\ff G]$ trivially
vanishes for 
$\ff U=0$ [Eq.\ (\ref{eq:bl})].
The functional derivative of $\widehat{\Phi}[\ff G]$ with respect to $\ff G$ corresponds
to the removal of a propagator from each of the $\Phi$ diagrams. 
Taking care of topological factors, \cite{LW60,AGD64} one ends up with the skeleton-diagram 
expansion for the self-energy, i.e.\ one gets Eq.\ (\ref{eq:der}).
Using Eq.\ (\ref{eq:der}), the Dyson equation (\ref{eq:dyson}), and 
$\Phi_{\ff t, \ff U} \equiv \widehat{\Phi}_{\ff U}[\ff G_{\ff t, \ff U}]$, the $\mu$
derivative of the l.h.s and of the r.h.s of Eq.\ (\ref{eq:om}) are equal for any fixed 
interaction strength $\ff U$ and temperature $T$. Namely,
$(\partial / \partial \mu) (\Phi_{\ff t, \ff U} + \Tr \ln \ff G_{\ff t, \ff U} 
- \Tr \, \ff \Sigma_{\ff t, \ff U} \ff G_{\ff t, \ff U}) =  
\Tr \, \ff G_{\ff t, \ff U}^{-1} (\partial \ff G_{\ff t, \ff U} / \partial \mu) 
- \Tr \, \ff G_{\ff t, \ff U} (\partial \ff \Sigma_{\ff t, \ff U} / 
\partial \mu) = - \Tr \: \ff G_{\ff t, \ff U} = - \langle N \rangle_{\ff t, \ff U} 
= \partial \Omega_{\ff t, \ff U} / \partial \mu$. 
Integration over $\mu$ then yields Eq.\ (\ref{eq:om}).
(Note that Eq.\ (\ref{eq:om}) holds trivially for $\mu \to -\infty$, i.e.\ for 
$\langle N \rangle_{\ff t, \ff U} \to 0$ since $\ff \Sigma_{\ff t, \ff U} = 0$ 
and $\Phi_{\ff t, \ff U} =0$ in this limit).
An equivalent derivation of Eq.\ (\ref{eq:om}) can be given by a coupling-constant
integration. \cite{LW60}

\section{Luttinger-Ward functional}
\label{sec:lw}

The starting point is the standard functional-integral representation of the 
partition function as given in Ref.\ \onlinecite{NO88}, for example:
Define the functional
\be
   \widehat{\Omega}_{\ff U}[\ff G_0^{-1}] = 
   - T \ln \widehat{Z}_{\ff U}[\ff G_0^{-1}]
\labeq{omgfree}
\ee
with
\be
   \widehat{Z}_{\ff U}[\ff G_0^{-1}] = \int D \xi D \xi^\ast 
   \exp\left( \widehat{A}_{\ff U,\xi\xi^\ast}[\ff G_0^{-1}] \right)
\ee
and
\ba
   &&
   \widehat{A}_{\ff U,\xi\xi^\ast}[\ff G_0^{-1}] =
   \sum_{n,\alpha\beta} \xi_\alpha^\ast(i\omega_n) 
   G_{0,\alpha\beta}^{-1}(i\omega_n)
   \xi_\beta(i\omega_n) 
\nonumber \\   
   &&- 
   \frac{1}{2} \sum_{\alpha\beta\gamma\delta} U_{\alpha\beta\delta\gamma} \int_0^{1/T} 
   \!\! d\tau \:
   \xi_\alpha^\ast(\tau) 
   \xi_\beta^\ast(\tau)
   \xi_\gamma(\tau) 
   \xi_\delta(\tau) \: .
\ea
$\widehat{\Omega}_{\ff U}[\ff G_0^{-1}]$ parametrically depends on $\ff U$.
$\ff G_0^{-1}$ is considered as a free variable.
At the (matrix inverse of the) exact free Green's function, $\ff G_0^{-1} = 
\ff G_{\ff t, 0}^{-1}$, the functional yields the exact grand potential,
\be
  \widehat{\Omega}_{\ff U}[\ff G_{\ff t, 0}^{-1}] = \Omega_{\ff t, \ff U} \: ,
\labeq{omex0} 
\ee
of the system with Hamiltonian $H = H_0(\ff t) + H_{1}(\ff U)$.
Its derivative defines a functional
$\widehat{\cal \ff G}_{\ff U}[\ff G_0^{-1} ]$,
\be
\frac{1}{T} \frac{\delta \widehat{\Omega}_{\ff U}[\ff G_0^{-1}]}
{\delta \ff G_0^{-1}} 
= - \: \frac{1}{\widehat{Z}_{\ff U}[\ff G_0^{-1}]} 
\frac{\delta \widehat{Z}_{\ff U}[\ff G_0^{-1}]}
{\delta \ff G_0^{-1}} 
\equiv - \widehat{\cal \ff G}_{\ff U}[\ff G_0^{-1}] \: ,
\labeq{omder}
\ee
with the property
\be
  \widehat{\cal \ff G}_{\ff U}[\ff G_{\ff t, 0}^{-1}] = \ff G_{\ff t, \ff U} 
\labeq{calg}
\ee
which is easily verified using Eq.\ (\ref{eq:gfunct}).

The strategy to be pursued is the following: 
$\widehat{\cal \ff G}_{\ff U}[\ff G_{0}^{-1}]$ is a universal ($\ff t$ independent) 
functional and can be used to construct a universal relation 
$\ff G = \widehat{\ff G}_{\ff U}[\ff \Sigma]$ 
between the one-particle Green's function and the self-energy independent
from the Dyson equation. 
Using the universal functionals $\widehat{\Omega}_{\ff U}[\ff G_{0}^{-1}]$ and 
$\widehat{\ff G}_{\ff U}[\ff \Sigma]$, a universal functional 
$\widehat{F}_{\ff U}[\ff \Sigma]$ is defined the derivative of which essentially 
yields $\widehat{\ff G}_{\ff U}[\ff \Sigma]$.
The Luttinger-Ward functional can then be obtained by Legendre transformation
and is universal by construction.

To start with, consider the equation
\be
  \widehat{\cal \ff G}_{\ff U}[ \ff G^{-1} + \ff \Sigma ] = \ff G \: .
\labeq{rel}  
\ee
This is a relation between the variables $\ff G$ and $\ff \Sigma$ which,
for a given $\ff \Sigma$, may be solved for $\ff G$. 
This defines a functional $\widehat{\ff G}_{\ff U}[\ff \Sigma]$, i.e.
\be
  \widehat{\cal \ff G}_{\ff U}[ \widehat{\ff G}_{\ff U}[\ff \Sigma]^{-1} + \ff \Sigma ] = 
  \widehat{\ff G}_{\ff U}[\ff \Sigma] \: .
\labeq{gfunc}
\ee
For a given self-energy $\ff \Sigma$, the Green's function 
$\ff G = \widehat{\ff G}_{\ff U}[\ff \Sigma]$ is defined to be the solution of
Eq.\ (\ref{eq:rel}).
From the Dyson equation (\ref{eq:dyson}) and Eq.\ (\ref{eq:calg}) it is obvious
that the relation (\ref{eq:rel}) is satisfied for $\ff G$ and $\ff \Sigma$
being the exact Green's function and the exact self-energy,
$\ff G=\ff G_{\ff t, \ff U}$ and $\ff \Sigma=\ff \Sigma_{\ff t, \ff U}$, of a system 
with the interaction $\ff U$ and some set of one-particle parameters $\ff t$ 
($H = H_0(\ff t) + H_{1}(\ff U)$). 
Hence,
\be
  \widehat{\ff G}_{\ff U}[ \ff \Sigma_{\ff t, \ff U}] = 
  \ff G_{\ff t, \ff U} \: .
\labeq{gex}
\ee
A brief discussion of the existence and the uniqueness of possible solutions 
of the relation (\ref{eq:rel}) is given in Appendix \ref{sec:a}.

With the help of the functionals $\widehat{\Omega}_{\ff U}[\ff G_0^{-1}]$ and 
$\widehat{\ff G}_{\ff U}[\ff \Sigma]$, a functional $\widehat{F}_{\ff U}[\ff \Sigma]$ 
can be defined as:
\be
  \widehat{F}_{\ff U}[\ff \Sigma] = 
  \widehat{\Omega}_{\ff U}[ 
  \widehat{\ff G}_{\ff U}[\ff \Sigma]^{-1} + \ff \Sigma
  ]
  - 
  \Tr \ln \widehat{\ff G}_{\ff U}[\ff \Sigma] \: .
\labeq{fdef}
\ee
Using \refeq{omder} one finds: 
\ba
\frac{1}{T} \frac{\delta \widehat{F}_{\ff U}[\ff \Sigma]}{\delta \ff \Sigma} 
  &=&
- \widehat{\cal \ff G}_{\ff U}[ 
  \widehat{\ff G}_{\ff U}[\ff \Sigma]^{-1} + \ff \Sigma
  ] \cdot
  \left(
  \frac{\delta \widehat{\ff G}_{\ff U}[\ff \Sigma]^{-1}}{\delta \ff \Sigma} + \ff 1
  \right)
\nonumber \\ 
  &-& \widehat{\ff G}_{\ff U}[\ff \Sigma]^{-1} \cdot
  \frac{\delta \widehat{\ff G}_{\ff U}[\ff \Sigma]}{\delta \ff \Sigma} \: ,
\ea
and, using \refeq{gfunc},
\be
\frac{1}{T} \frac{\delta \widehat{F}_{\ff U}[\ff \Sigma]}{\delta \ff \Sigma} 
=
- \widehat{\ff G}_{\ff U}[\ff \Sigma] \: .
\label{eq:fder}
\ee
So $\widehat{\ff G}_{\ff U}[\ff \Sigma]$ can be considered as the gradient of the
(scalar) self-energy functional $\widehat{F}_{\ff U}[\ff \Sigma]$.
Therewith, the Legendre transform of $\widehat{F}_{\ff U}[\ff \Sigma]$ can be
constructed:
\be
  \widehat{\Phi}_{\ff U}[\ff G] = \widehat{F}_{\ff U}[\widehat{\ff \Sigma}_{\ff U}[\ff G]] + 
  \Tr ( \widehat{\ff \Sigma}_{\ff U}[\ff G] \; \ff G ) \: .
\label{eq:llww}
\ee
Here $\widehat{\ff \Sigma}_{\ff U}[\ff G]$ is the inverse of the functional 
$\widehat{\ff G}_{\ff U}[\ff \Sigma]$.
The functional can be assumed to be invertible (locally) provided that the system is not 
at a critical point for a phase transition (see also Ref.\ \onlinecite{Pot03a}). 
Eq.\ (\ref{eq:llww}) defines the Luttinger-Ward functional.

\section{Discussion}
\label{sec:dis}

\subsection{Properties of the Luttinger-Ward functional}

The properties of the Luttinger-Ward functional, Eqs.\ (\ref{eq:om}--\ref{eq:univ}),
can be verified easily:
Eqs.\ (\ref{eq:dyson}), (\ref{eq:omex0}), (\ref{eq:gex}) and (\ref{eq:fdef}) imply 
\be
\widehat{F}_{\ff U}[\ff \Sigma_{\ff t,\ff U}] =
{\Omega}_{\ff t,\ff U} - \Tr \ln {\ff G}_{\ff t, \ff U} \: ,
\label{eq:fex}
\ee
and with
$\widehat{\ff \Sigma}_{\ff U}[\ff G_{\ff t,\ff U}] = \ff \Sigma_{\ff t,\ff U}$
the evaluation of the Luttinger-Ward functional at $\ff G = \ff G_{\ff t,\ff U}$
yields 
\be
\Phi_{\ff t,\ff U} \equiv \widehat{\Phi}_{\ff U}[\ff G_{\ff t,\ff U}] =
{\Omega}_{\ff t,\ff U} - \Tr \ln {\ff G}_{\ff t, \ff U} + \Tr ( \ff \Sigma_{\ff t,\ff U}
{\ff G}_{\ff t, \ff U}) \: ,
\label{eq:phieval}
\ee
i.e.\ Eq.\ (\ref{eq:om}).
From Eqs.\ (\ref{eq:fder}) and (\ref{eq:llww}), one immediately has:
\be
\frac{1}{T} \frac{\delta \widehat{\Phi}_{\ff U}[\ff G]}{\delta \ff G} 
=
\widehat{\ff \Sigma}_{\ff U}[\ff G] \: ,
\label{eq:llwwder}
\ee
i.e.\ Eq.\ (\ref{eq:der}). 
In the limit $\ff U = 0$, the functionals 
$\widehat{\ff G}_{\ff U= 0}[\ff \Sigma]$ and $\widehat{F}_{\ff U= 0}[\ff \Sigma]$
are ill-defined (the domain of the functionals shrinks to a single point, 
$\ff \Sigma = 0$, see Appendix \ref{sec:a}). 
However, from Eq.\ (\ref{eq:phieval}), one directly has
$\Phi_{\ff U = 0}[\ff G_{\ff t,0}] = 0$ for any $\ff t$ 
[see Eq.\ (\ref{eq:bl})] since
$\ff \Sigma_{\ff t,\ff 0} = 0$ and 
${\Omega}_{\ff t,\ff 0} = \Tr \ln {\ff G}_{\ff t, \ff 0}$
(a proof for the latter can be found in Ref.\ \onlinecite{LW60}).
Finally, the universality of $\widehat{\Phi}_{\ff U}[\ff G]$, Eq.\ (\ref{eq:univ}) 
is obvious
as the definition (\ref{eq:llww}) of the Luttinger-Ward functional involves
the universal ($\ff t$ independent) functionals
$\widehat{F}_{\ff U}[\ff \Sigma]$ and $\widehat{\ff \Sigma}_{\ff U}[\ff G]$
only.

\subsection{Variational principle}

Using the Legendre transform of the Luttinger-Ward functional 
$\widehat{F}_{\ff U}[\ff \Sigma]$, one may define
\be
  \widehat{\Omega}_{\ff t, \ff U}[\ff \Sigma] = 
  \Tr \ln \frac{1}{\ff G_{\ff t,0}^{-1} - \ff \Sigma}
  + \widehat{F}_{\ff U}[\ff \Sigma] \: .
\labeq{sef}
\ee
The functional derivative is easily calculated:
\be
  \frac{1}{T} \frac{\delta \widehat{\Omega}_{\ff t, \ff U}[\ff \Sigma]}
  {\delta \ff \Sigma} = 
  \frac{1}{\ff G_{\ff t,0}^{-1} - \ff \Sigma} - 
  \widehat{\ff G}_{\ff U}[\ff \Sigma] \: .
\ee
The equation 
\be
  \widehat{\ff G}_{\ff U}[\ff \Sigma] = \frac{1}{\ff G_{\ff t,0}^{-1} - \ff \Sigma}
\label{eq:sig}
\ee
is a (highly non-linear) conditional equation for the self-energy of the system 
$H = H_0(\ff t) + H_{1}(\ff U)$:
Eqs.\ (\ref{eq:dyson}) and (\ref{eq:gex}) show that it is satisfied by the exact 
self-energy $\ff \Sigma = \ff \Sigma_{\ff t, \ff U}$.
Note that the l.h.s of (\ref{eq:sig}) is independent of $\ff t$ but depends 
on $\ff U$ (universality of $\widehat{\ff G}[\ff \Sigma]$), while the r.h.s  
is independent of $\ff U$ but depends on $\ff t$ via $\ff G_{\ff t,0}^{-1}$.
The obvious problem of finding a solution of \refeq{sig} is that there is no
closed form for the functional $\widehat{\ff G}_{\ff U}[\ff \Sigma]$.
Solving Eq.\ (\ref{eq:sig}) is equivalent to a search for the stationary point 
of the grand potential as a functional of the self-energy:
\be
  \frac{\delta \widehat{\Omega}_{\ff t, \ff U}[\ff \Sigma]}
  {\delta \ff \Sigma} = 0 \; .
\label{eq:var}
\ee
Similarly, one can also construct a variational principle using the Green's function
as the basic variable, $\delta \widehat{\Omega}_{\ff t, \ff U}[\ff G] / \delta \ff G = 0$.

\subsection{Dynamical mean-field theory}

The dynamical mean-field theory \cite{MV89,GK92a,Jar92,GKKR96} basically applies
to lattice models of correlated electrons with on-site interactions such as the
Hubbard model, \cite{Hub63} for example. The DMFT aims at an approximate solution
of Eq.\ (\ref{eq:sig}) and is based on two ingredients: 

(i) It is important to note that the Luttinger-Ward functional 
$\widehat{\Phi}_{\ff U}[\ff G]$ is the same for the lattice
(e.g.\ Hubbard) model and for an impurity model (single-impurity Anderson model).
Actually a (decoupled) set of impurity models has to be considered -- one impurity model  
with the according local interaction at each site of the original lattice.
This ensures that the interaction ($\ff U$) term is the same as in the lattice model.
(In case of translational symmetry the a priori different impurity models can be 
assumed to be equivalent).
As $\ff U$ is the same in the lattice and in the impurity model, the Luttinger-Ward
functional, as well as $\widehat{\ff G}_{\ff U}[\ff \Sigma]$, is the same.

(ii) Let the lattice model be characterized by one-particle parameters $\ff t$ and the 
impurity model by parameters $\ff t'$.
The fundamental equation (\ref{eq:sig}) for the lattice model would then be solved by the exact 
self-energy $\ff \Sigma_{\ff t, \ff U}$.
As an ansatz for an approximate solution $\ff \Sigma$ of Eq.\ (\ref{eq:sig}), the self-energy 
is assumed to be local within the DMFT and to be representable as the exact self-energy of 
the impurity model for some parameters $\ff t'$:
\be
  \ff \Sigma = \ff \Sigma_{\ff t', \ff U} \: .
\label{eq:an}
\ee

The universality of the Luttinger-Ward functional (i) and the local approximation for the
self-energy (ii) are sufficient to derive the DMFT:
Inserting the ansatz (\ref{eq:an}) into Eq.\ (\ref{eq:sig}) yields a conditional equation 
for the one-particle parameters of the impurity model $\ff t'$.
The l.h.s becomes 
$\widehat{\ff G}_{\ff U}[\ff \Sigma_{\ff t', \ff U}]
= {\ff G}_{\ff t',\ff U}$, i.e.\ the exact Green's function of the impurity model,
while the r.h.s reads $(\ff G_{\ff t,0}^{-1} - \ff \Sigma_{\ff t', \ff U})^{-1}$.
The resulting equation for the parameters $\ff t'$ can be fulfilled only locally,
i.e.\ by equating the local elements of the respective Green's functions at the
impurity and the original site respectively:
\be
  \left( {\ff G}_{\ff t',\ff U} \right)_{\rm loc}
  = 
  \left( \frac{1}{\ff G_{\ff t,0}^{-1} - \ff \Sigma_{\ff t', \ff U}} \right)_{\rm loc} \: .
\label{eq:sc}
\ee
This is the so-called self-consistency equation of the DMFT. \cite{GKKR96}

This consideration can be seen as an independent and, in particular, non-perturbative
re-derivation of the DMFT which supplements known approaches such as the cavity method. 
\cite{GKKR96}

\subsection{Self-energy-functional approach}

The universality of the Luttinger-Ward functional or of its Legendre transform
$\widehat{F}_{\ff U}[\ff \Sigma]$ is central to the recently developed
self-energy-functional approach. \cite{Pot03a,Pot03b}
The SFA is a general variational scheme which includes the DMFT as a special
limit.
The idea is to take as an ansatz for the self-energy of a model
$H = H_0(\ff t) + H_{1}(\ff U)$ the exact self-energy $\ff \Sigma_{\ff t', \ff U}$
of a so-called reference system $H' = H_0(\ff t') + H_{1}(\ff U)$ that shares with
the original model the same interaction part.
The parameters $\ff t'$ of the one-particle part are considered as variational 
parameters to search for the stationary point of the grand potential as a 
functional of the self-energy.
This means to insert the ansatz $\ff \Sigma = \ff \Sigma_{\ff t', \ff U}$ into the
general expression (\ref{eq:sef}) and to solve the Euler equation
$\partial \widehat{\Omega}_{\ff t, \ff U}[ \ff \Sigma_{\ff t', \ff U}] / \partial \ff t' = 0$,
i.e.:
\be
  \frac{\partial}{\partial \ff t'} \left( 
  \Tr \ln \frac{1}{\ff G_{\ff t,0}^{-1} - \ff \Sigma_{\ff t', \ff U}}
  + \widehat{F}_{\ff U}[\ff \Sigma_{\ff t', \ff U}] 
  \right) = 0
\labeq{eu}
\ee
for $\ff t'$.
If the search for the optimum set of one-particle parameters $\ff t'$ was unrestricted,
the approach would be exact in principle as the Euler equation would then be equivalent
with the Euler equation (\ref{eq:sig}) of the general variational principle Eq.\ (\ref{eq:var}).

A restriction of the space of variational parameters becomes necessary to evaluate the 
quantity $\widehat{\Omega}_{\ff t, \ff U}[ \ff \Sigma_{\ff t', \ff U}]$ which, in general, 
is impossible as a closed form for the functional $\widehat{F}_{\ff U}[\ff \Sigma]$ is not 
known.
With a proper restriction, however, the reference system $H'$ can be made accessible to an 
exact (numerical) solution which allows to derive the exact grand potential and the exact 
Green's function of the system $H'$.
Therewith, making use of the universality of $\widehat{F}_{\ff U}[\ff \Sigma]$ and using 
Eqs.\ (\ref{eq:llww}) and (\ref{eq:phieval}) for the reference system,
\be
  \widehat{F}_{\ff U}[\ff \Sigma_{\ff t', \ff U}] = {\Omega}_{\ff t', \ff U}
  -
  \Tr \ln \ff G_{\ff t',\ff U} \: .
\ee
Note that this implies that an exact evaluation of $\widehat{F}_{\ff U}[\ff \Sigma]$
is possible for self-energies of a exactly solvable reference system with the same
interaction part as the original one.
Using this result in Eq.\ (\ref{eq:eu}), one obtains:
\be
  \frac{\partial}{\partial \ff t'} \left( 
  {\Omega}_{\ff t', \ff U}
  +
  \Tr \ln \frac{1}{\ff G_{\ff t,0}^{-1} - \ff \Sigma_{\ff t', \ff U}}
  -
  \Tr \ln \ff G_{\ff t',\ff U} 
  \right) = 0 \: ,
\labeq{eus}
\ee
which can be evaluated to fix $\ff t'$ and therewith the optimal self-energy
and grand potential (see Refs.\ \onlinecite{Pot03a,Pot03b,PAD03,DAH+04} for details
and concrete examples).

\subsection{Luttinger's theorem}

Finally, the role of the Luttinger-Ward functional in the derivation of general
properties of correlated electron systems shall be discussed. 
As an important example, the Luttinger theorem \cite{LW60} is considered.
For a translationally invariant system, the theorem states that in the limit
$T \to 0$ the average 
particle number is equal to the volume enclosed by the Fermi surface in 
$\ff k$ space:
\be
  \langle N \rangle = V_{\rm FS} \: .
\label{eq:lt}
\ee
The Fermi surface is defined by the set of $\ff k$ points in the first Brillouin
zone that satisfies $\mu - \eta_{\ff k} = 0$ where $\eta_{\ff k}$ are the eigenvalues 
of the matrix $\ff t + \ff \Sigma(\omega)$ at vanishing excitation energy $\omega=0$.
Hence, to formulate the Luttinger theorem, one obviously has to presuppose that there 
is a Fermi surface at all, i.e.\ that $\ff \Sigma(\omega=0)$ is Hermitian.
\footnote{
For systems without Fermi surface, there is no Luttinger theorem.
A nice example is given by the Falicov-Kimball model in infinite 
dimensions, see Ref.\ \onlinecite{FZ03}.
}
The original proof of the theorem \cite{LW60} is perturbative as it makes use
of the skeleton-diagram expansion. 
A non-perturbative proof, based on topological considerations, was proposed
recently \cite{Osh00} and is based on the assumption that the system is a 
Fermi liquid.

To discuss the Luttinger theorem in the present context, consider the following
shift transformation of the Green's function
\be
  \ff S^{(z)} \ff G(i\omega_n) = \ff G^{(z)}(i\omega_n) = \ff G(i\omega_n+iz)
\ee
with $z = 2 \pi k T$ and $k$ integer ($z$ is a bosonic Matsubara frequency).
$\ff S^{(z)}$ is a linear and unitary transformation.
The shift transformation leaves the functional integral Eq.\ (\ref{eq:omgfree}) unchanged:
\be
  \widehat{\Omega}_{\ff U}[\ff S^{(z)} \ff G_0^{-1}]
  =
  \widehat{\Omega}_{\ff U}[\ff G_0^{-1}] \: .
\label{eq:invar}
\ee
To verify this invariance, one has to note that the shift of the Matsubara frequencies 
in $\ff G_0^{-1}$ by $z$ can be transformed into a shift $\omega_n \to \omega_n - z$ 
in the Grassmann numbers:
\be 
  \xi_\alpha(i\omega_n) \to \xi_\alpha (i\omega_n-iz) \: .
\labeq{tr1}  
\ee
In imaginary-time representation this shift is equivalent with the 
multiplication of a phase:
\be
  \xi_\alpha(\tau) \to e^{-iz \tau} \xi_\alpha(\tau) \; , \qquad
  \xi_\alpha^\ast(\tau) \to e^{iz \tau} \xi_\alpha^\ast(\tau) \: . 
\labeq{tr2}  
\ee
This, however, leaves the functional integral unchanged as the transformation 
\refeq{tr1} or \refeq{tr2} is linear and the Jacobian is unity.
Note that antiperiodic boundary conditions $\xi_\alpha(\tau=1/T)=-\xi_\alpha(\tau=0)$
are respected for a bosonic shift frequency $z$.

Denoting 
$\Omega_{\ff t, \ff U}(z) \equiv \widehat{\Omega}_{\ff U}[\ff S^{(z)} \ff G_{\ff t,0}^{-1}]$,
Eq.\ (\ref{eq:invar}) states that $\Omega_{\ff t, \ff U}(z)=\Omega_{\ff t, \ff U}(0)$.
Following the steps in the construction of the Luttinger-Ward functional in Sec.\ \ref{sec:lw},
one easily verifies that this implies $\Phi_{\ff t, \ff U}(z)=\Phi_{\ff t, \ff U}(0)$ where
$\Phi_{\ff t, \ff U}(z) \equiv \widehat{\Phi}_{\ff U}[\ff S^{(z)} \ff G_{\ff t, \ff U}]$.
For the Legendre transform, one has $F_{\ff t, \ff U}(z)=F_{\ff t, \ff U}(0)$ where
$F_{\ff t, \ff U}(z) \equiv \widehat{F}_{\ff U}[\ff S^{(z)} \ff \Sigma_{\ff t, \ff U}]$.
Now, in the limit $T \to 0$, $z$ becomes a continuous variable.
Hence,
\be
  \frac{d}{dz} \lim_{T \to 0} F_{\ff t, \ff U}(z) = 0 \: .
\label{eq:ddzf}
\ee
If the limit and the derivative can be interchanged,
\be
   \frac{d}{dz} \lim_{T \to 0} F_{\ff t, \ff U}(z) 
   =
   \lim_{T \to 0} \frac{d}{dz} F_{\ff t, \ff U}(z) \: ,
\label{eq:cond}      
\ee
Eqs.\ (\ref{eq:fex}) and (\ref{eq:ddzf}) imply
\be
  - \lim_{T \to 0} \frac{d\Omega_{\ff t, \ff U}(z)}{dz} 
  =
  - \lim_{T \to 0} \frac{d \Tr \ln \ff S^{(z)} \ff G_{\ff t, \ff U}}{dz} \: .
\label{eq:x}  
\ee
The $z$ dependence of the grand potential is the same as its $\mu$ dependence, 
and thus $ - (d/dz) \Omega_{\ff t, \ff U}(z=0) = - \partial \Omega_{\ff t, \ff U} / \partial \mu 
= \langle N \rangle$.
The evaluation of the r.h.s in Eq.\ (\ref{eq:x}) is straightforward and can be found 
in Ref.\ \onlinecite{LW60}, for example.
It turns out that at $z=0$ the r.h.s is just the Fermi-surface volume $V_{\rm FS}$.

Consequently, the non-perturbative construction of the Luttinger-Ward functional allows 
to reduce the proof of the Luttinger theorem to the proof of Eq.\ (\ref{eq:cond}).
This, however, requires certain assumptions on the regularity of the $T \to 0$ limit
which are non-trivial generally.

\section{Summary}
\label{sec:sum}

To summarize, the present paper has shown that the Luttinger-Ward functional can
be constructed within the framework of functional integrals under fairly general
assumptions. 
In particular, there no need for an adiabatic connection to the non-interacting
limit and no expansion in the interaction strength as was required in the original
approach of Luttinger and Ward. \cite{LW60}
The construction merely assumes the very existence of the functional integral over
Grassmann fields, i.e.\ the existence of the Trotter limit, for the representation
of the partition function.

It is well known that the Luttinger-Ward functional can be employed for different 
purposes, some of which have been discussed here:
The functional is used to derive some general properties of correlated electron systems,
such as the Luttinger theorem.
It allows to formulate a variational principle involving a thermodynamical potential
as a functional of the Green's function or the self-energy and thereby provides
a unique and thermodynamically meaningful link 
between static and dynamic quantities which is helpful for interpretations and for 
the construction of approximations.
An independent derivation of the dynamical mean-field is possible using the special 
properties of the Luttinger-Ward functional and the universality of the functional 
in particular.
The latter is of central importance in the context of the self-energy-functional approach
which is a general framework to construct thermodynamically consistent approximations.

Referring to the standard definition of the Luttinger-Ward functional that is based
on the weak-coupling skeleton-diagram expansion, the above-mentioned and any further 
considerations based on the functional and its unique properties meet with criticism 
when applied to strongly correlated, non-Fermi liquid or symmetry-broken states. 
This is exactly the point where the presented non-perturbative construction of the
Luttinger-Ward functional is useful.

\acknowledgments
The author acknowledges helpful discussions with F.F.\ Assaad, M.\ Bechmann, 
W.\ Hanke and G.\ Kotliar.
The work is supported by the Deutsche Forschungsgemeinschaft (Forschergruppe
538).

\appendix

\section{}
\label{sec:a}

As the relation (\ref{eq:rel}) is highly non-linear, the existence and the uniqueness
of possible solutions have to be discussed:

Take $\ff U$ to be fixed and assume that the self-energy given is the exact self-energy 
of a system $H = H_0(\ff t) + H_{1}(\ff U)$ with some hopping parameters $\ff t$.
So the self-energy $\ff \Sigma$ is assumed to be given from the space ${\cal S}_{\ff U}$
of $\ff t$ representable self-energies ${\cal S}_{\ff U} \equiv \{ \ff \Sigma \, | \, \ff \Sigma 
= \ff \Sigma_{\ff t, \ff U} \, , \, \ff t \: \mbox{arbitrary} \}$
($\ff U$ fixed).
With the help of Eq.\ (\ref{eq:calg}) it is then obvious that the exact Green's function 
of this system, $\ff G = \ff G_{\ff t, \ff U}$, solves Eq.\ (\ref{eq:rel}) as the Dyson 
equation (\ref{eq:dyson}) shows that 
$\ff G_{\ff t, \ff U}^{-1} + \ff \Sigma_{\ff t, \ff U}$ is the exact free Green's
function of this system.
Concluding, one has 
$\widehat{\ff G}_{\ff U}[ \ff \Sigma_{\ff t, \ff U}] = \ff G_{\ff t, \ff U}$,
and thus the existence of a solution is guaranteed on the space ${\cal S}_{\ff U}$.
Note that it is very convenient to consider ${\cal S}_{\ff U}$ as the domain of 
the functional $\widehat{\ff G}_{\ff U}[\ff \Sigma]$ since this ensures the correct
analytical and causal properties of the variable $\ff \Sigma$.

Under the functional $\widehat{\ff G}_{\ff U}[\ff \Sigma]$ the space ${\cal S}_{\ff U}$
is mapped onto the space ${\cal G}_{\ff U}$
of $\ff t$ representable Green's functions ${\cal G}_{\ff U} \equiv \{ \ff G \, | \, \ff G
= \ff G_{\ff t, \ff U} \, , \, \ff t \: \mbox{arbitrary} \}$
($\ff U$ fixed).
Generally, the map $\widehat{\ff G}_{\ff U} : {\cal S}_{\ff U} \to {\cal G}_{\ff U}$
is not unique. \cite{Pot03a}
Hence, the uniqueness of the functional $\widehat{\ff G}_{\ff U}[\ff \Sigma]$ must be
enforced by a proper restriction of the range ${\cal G}_{\ff U}$, i.e.\ of
the solution set of Eq.\ (\ref{eq:rel}). 
The considerations in Secs.\ \ref{sec:lw} and \ref{sec:dis}, however, are unaffected 
and hold for any choice of the range, 
see also the related discussion in Ref.\ \onlinecite{Pot03a}.


\begin{thebibliography}{24}
\expandafter\ifx\csname natexlab\endcsname\relax\def\natexlab#1{#1}\fi
\expandafter\ifx\csname bibnamefont\endcsname\relax
  \def\bibnamefont#1{#1}\fi
\expandafter\ifx\csname bibfnamefont\endcsname\relax
  \def\bibfnamefont#1{#1}\fi
\expandafter\ifx\csname citenamefont\endcsname\relax
  \def\citenamefont#1{#1}\fi
\expandafter\ifx\csname url\endcsname\relax
  \def\url#1{\texttt{#1}}\fi
\expandafter\ifx\csname urlprefix\endcsname\relax\def\urlprefix{URL }\fi
\providecommand{\bibinfo}[2]{#2}
\providecommand{\eprint}[2][]{\url{#2}}

\bibitem[{\citenamefont{Abrikosow et~al.}(1964)\citenamefont{Abrikosow, Gorkov,
  and Dzyaloshinski}}]{AGD64}
\bibinfo{author}{\bibfnamefont{A.~A.} \bibnamefont{Abrikosow}},
  \bibinfo{author}{\bibfnamefont{L.~P.} \bibnamefont{Gorkov}},
  \bibnamefont{and} \bibinfo{author}{\bibfnamefont{I.~E.}
  \bibnamefont{Dzyaloshinski}}, \emph{\bibinfo{title}{Methods of Quantum Field
  Theory in Statistical Physics}} (\bibinfo{publisher}{Prentice-Hall},
  \bibinfo{address}{New Jersey}, \bibinfo{year}{1964}).

\bibitem[{\citenamefont{Fetter and Walecka}(1971)}]{FW71}
\bibinfo{author}{\bibfnamefont{A.~L.} \bibnamefont{Fetter}} \bibnamefont{and}
  \bibinfo{author}{\bibfnamefont{J.~D.} \bibnamefont{Walecka}},
  \emph{\bibinfo{title}{Quantum Theory of Many-Particle Systems}}
  (\bibinfo{publisher}{McGraw-Hill}, \bibinfo{address}{New York},
  \bibinfo{year}{1971}).

\bibitem[{\citenamefont{Negele and Orland}(1988)}]{NO88}
\bibinfo{author}{\bibfnamefont{J.~W.} \bibnamefont{Negele}} \bibnamefont{and}
  \bibinfo{author}{\bibfnamefont{H.}~\bibnamefont{Orland}},
  \emph{\bibinfo{title}{Quantum Many-Particle Systems}}
  (\bibinfo{publisher}{Addison-Wesley}, \bibinfo{address}{Redwood City},
  \bibinfo{year}{1988}).

\bibitem[{\citenamefont{Luttinger and Ward}(1960)}]{LW60}
\bibinfo{author}{\bibfnamefont{J.~M.} \bibnamefont{Luttinger}}
  \bibnamefont{and} \bibinfo{author}{\bibfnamefont{J.~C.} \bibnamefont{Ward}},
  \bibinfo{journal}{Phys. Rev.} \textbf{\bibinfo{volume}{118}},
  \bibinfo{pages}{1417} (\bibinfo{year}{1960}).

\bibitem[{\citenamefont{Hubbard}(1963)}]{Hub63}
\bibinfo{author}{\bibfnamefont{J.}~\bibnamefont{Hubbard}},
  \bibinfo{journal}{Proc. R. Soc. London A} \textbf{\bibinfo{volume}{276}},
  \bibinfo{pages}{238} (\bibinfo{year}{1963}).

\bibitem[{\citenamefont{Baym}(1962)}]{Bay62}
\bibinfo{author}{\bibfnamefont{G.}~\bibnamefont{Baym}}, \bibinfo{journal}{Phys.
  Rev.} \textbf{\bibinfo{volume}{127}}, \bibinfo{pages}{1391}
  (\bibinfo{year}{1962}).

\bibitem[{\citenamefont{Luttinger}(1960)}]{Lut60}
\bibinfo{author}{\bibfnamefont{J.~M.} \bibnamefont{Luttinger}},
  \bibinfo{journal}{Phys. Rev.} \textbf{\bibinfo{volume}{119}},
  \bibinfo{pages}{1153} (\bibinfo{year}{1960}).

\bibitem[{\citenamefont{Baym and Kadanoff}(1961)}]{BK61}
\bibinfo{author}{\bibfnamefont{G.}~\bibnamefont{Baym}} \bibnamefont{and}
  \bibinfo{author}{\bibfnamefont{L.~P.} \bibnamefont{Kadanoff}},
  \bibinfo{journal}{Phys. Rev.} \textbf{\bibinfo{volume}{124}},
  \bibinfo{pages}{287} (\bibinfo{year}{1961}).

\bibitem[{\citenamefont{Metzner and Vollhardt}(1989)}]{MV89}
\bibinfo{author}{\bibfnamefont{W.}~\bibnamefont{Metzner}} \bibnamefont{and}
  \bibinfo{author}{\bibfnamefont{D.}~\bibnamefont{Vollhardt}},
  \bibinfo{journal}{Phys. Rev. Lett.} \textbf{\bibinfo{volume}{62}},
  \bibinfo{pages}{324} (\bibinfo{year}{1989}).

\bibitem[{\citenamefont{Georges and Kotliar}(1992)}]{GK92a}
\bibinfo{author}{\bibfnamefont{A.}~\bibnamefont{Georges}} \bibnamefont{and}
  \bibinfo{author}{\bibfnamefont{G.}~\bibnamefont{Kotliar}},
  \bibinfo{journal}{Phys. Rev. B} \textbf{\bibinfo{volume}{45}},
  \bibinfo{pages}{6479} (\bibinfo{year}{1992}).

\bibitem[{\citenamefont{Jarrell}(1992)}]{Jar92}
\bibinfo{author}{\bibfnamefont{M.}~\bibnamefont{Jarrell}},
  \bibinfo{journal}{Phys. Rev. Lett.} \textbf{\bibinfo{volume}{69}},
  \bibinfo{pages}{168} (\bibinfo{year}{1992}).

\bibitem[{\citenamefont{Georges et~al.}(1996)\citenamefont{Georges, Kotliar,
  Krauth, and Rozenberg}}]{GKKR96}
\bibinfo{author}{\bibfnamefont{A.}~\bibnamefont{Georges}},
  \bibinfo{author}{\bibfnamefont{G.}~\bibnamefont{Kotliar}},
  \bibinfo{author}{\bibfnamefont{W.}~\bibnamefont{Krauth}}, \bibnamefont{and}
  \bibinfo{author}{\bibfnamefont{M.~J.} \bibnamefont{Rozenberg}},
  \bibinfo{journal}{Rev. Mod. Phys.} \textbf{\bibinfo{volume}{68}},
  \bibinfo{pages}{13} (\bibinfo{year}{1996}).

\bibitem[{\citenamefont{Potthoff}(2003{\natexlab{a}})}]{Pot03a}
\bibinfo{author}{\bibfnamefont{M.}~\bibnamefont{Potthoff}},
  \bibinfo{journal}{Euro. Phys. J. B} \textbf{\bibinfo{volume}{32}},
  \bibinfo{pages}{429} (\bibinfo{year}{2003}{\natexlab{a}}).

\bibitem[{\citenamefont{Potthoff}(2003{\natexlab{b}})}]{Pot03b}
\bibinfo{author}{\bibfnamefont{M.}~\bibnamefont{Potthoff}},
  \bibinfo{journal}{Euro. Phys. J. B} \textbf{\bibinfo{volume}{36}},
  \bibinfo{pages}{335} (\bibinfo{year}{2003}{\natexlab{b}}).

\bibitem[{\citenamefont{Potthoff et~al.}(2003)\citenamefont{Potthoff, Aichhorn,
  and Dahnken}}]{PAD03}
\bibinfo{author}{\bibfnamefont{M.}~\bibnamefont{Potthoff}},
  \bibinfo{author}{\bibfnamefont{M.}~\bibnamefont{Aichhorn}}, \bibnamefont{and}
  \bibinfo{author}{\bibfnamefont{C.}~\bibnamefont{Dahnken}},
  \bibinfo{journal}{Phys. Rev. Lett.} \textbf{\bibinfo{volume}{91}},
  \bibinfo{pages}{206402} (\bibinfo{year}{2003}).

\bibitem[{\citenamefont{Dahnken et~al.}(2003)\citenamefont{Dahnken, Aichhorn,
  Hanke, Arrigoni, and Potthoff}}]{DAH+04}
\bibinfo{author}{\bibfnamefont{C.}~\bibnamefont{Dahnken}},
  \bibinfo{author}{\bibfnamefont{M.}~\bibnamefont{Aichhorn}},
  \bibinfo{author}{\bibfnamefont{W.}~\bibnamefont{Hanke}},
  \bibinfo{author}{\bibfnamefont{E.}~\bibnamefont{Arrigoni}}, \bibnamefont{and}
  \bibinfo{author}{\bibfnamefont{M.}~\bibnamefont{Potthoff}},
  \bibinfo{journal}{preprint cond-mat} \textbf{\bibinfo{volume}{0309407}}
  (\bibinfo{year}{2003}).

\bibitem[{\citenamefont{Hofstetter and Kehrein}(1999)}]{HK99}
\bibinfo{author}{\bibfnamefont{W.}~\bibnamefont{Hofstetter}} \bibnamefont{and}
  \bibinfo{author}{\bibfnamefont{S.}~\bibnamefont{Kehrein}},
  \bibinfo{journal}{Phys. Rev. B} \textbf{\bibinfo{volume}{59}},
  \bibinfo{pages}{R12732} (\bibinfo{year}{1999}).

\bibitem[{\citenamefont{Kehrein}(1998)}]{Keh98}
\bibinfo{author}{\bibfnamefont{S.}~\bibnamefont{Kehrein}},
  \bibinfo{journal}{Phys. Rev. Lett.} \textbf{\bibinfo{volume}{81}},
  \bibinfo{pages}{3912} (\bibinfo{year}{1998}).

\bibitem[{\citenamefont{Kotliar}(1999)}]{Kot99}
\bibinfo{author}{\bibfnamefont{G.}~\bibnamefont{Kotliar}},
  \bibinfo{journal}{Euro. Phys. J. B} \textbf{\bibinfo{volume}{11}},
  \bibinfo{pages}{27} (\bibinfo{year}{1999}).

\bibitem[{\citenamefont{Savrasov and Kotliar}(2001)}]{SK01}
\bibinfo{author}{\bibfnamefont{S.~Y.} \bibnamefont{Savrasov}} \bibnamefont{and}
  \bibinfo{author}{\bibfnamefont{G.}~\bibnamefont{Kotliar}},
  \bibinfo{journal}{preprint cond-mat} \textbf{\bibinfo{volume}{0106308}}
  (\bibinfo{year}{2001}).

\bibitem[{\citenamefont{Savrasov and Kotliar}(2003)}]{SK03}
\bibinfo{author}{\bibfnamefont{S.~Y.} \bibnamefont{Savrasov}} \bibnamefont{and}
  \bibinfo{author}{\bibfnamefont{G.}~\bibnamefont{Kotliar}},
  \bibinfo{journal}{preprint cond-mat} \textbf{\bibinfo{volume}{0308053}}
  (\bibinfo{year}{2003}).

\bibitem[{\citenamefont{Georges}(2004)}]{Geo04}
\bibinfo{author}{\bibfnamefont{A.}~\bibnamefont{Georges}},
  \bibinfo{journal}{preprint cond-mat} \textbf{\bibinfo{volume}{0403123}}
  (\bibinfo{year}{2004}).

\bibitem[{\citenamefont{Oshikawa}(2000)}]{Osh00}
\bibinfo{author}{\bibfnamefont{M.}~\bibnamefont{Oshikawa}},
  \bibinfo{journal}{Phys. Rev. Lett.} \textbf{\bibinfo{volume}{84}},
  \bibinfo{pages}{3370} (\bibinfo{year}{2000}).

\bibitem[{\citenamefont{Freericks and Zlati\'c}(2003)}]{FZ03}
\bibinfo{author}{\bibfnamefont{J.~K.} \bibnamefont{Freericks}}
  \bibnamefont{and} \bibinfo{author}{\bibfnamefont{V.}~\bibnamefont{Zlati\'c}},
  \bibinfo{journal}{Rev. Mod. Phys.} \textbf{\bibinfo{volume}{75}},
  \bibinfo{pages}{1333} (\bibinfo{year}{2003}).

\end{thebibliography}
\end{document}